\documentclass[letterpaper,english,aps,prl,twocolumn,showpacs,superscriptaddress,preprintnumbers]{revtex4}
\usepackage[T1]{fontenc}
\usepackage[latin9]{inputenc}
\usepackage{color}
\usepackage{amsmath}
\usepackage{amssymb}
\usepackage{esint}
\newcommand\+{\dagger}
\newcommand\dd{\partial}

\makeatletter

\pdfpageheight\paperheight
\pdfpagewidth\paperwidth

\@ifundefined{textcolor}{}
{%
 \definecolor{BLACK}{gray}{0}
 \definecolor{WHITE}{gray}{1}
 \definecolor{RED}{rgb}{1,0,0}
 \definecolor{GREEN}{rgb}{0,1,0}
 \definecolor{BLUE}{rgb}{0,0,1}
 \definecolor{CYAN}{cmyk}{1,0,0,0}
 \definecolor{MAGENTA}{cmyk}{0,1,0,0}
 \definecolor{YELLOW}{cmyk}{0,0,1,0}
}

\usepackage{hyperref}

\usepackage[usenames,dvipsnames,svgnames,table]{xcolor}

\makeatother

\usepackage{babel}
\begin{document}

\preprint{EFI 14-41}
\title{Lowest Landau Level Stress Tensor and
  Structure Factor of\\ Trial Quantum Hall Wave Functions}

\author{Dung Xuan Nguyen}
\email{nxdung86@uchicago.edu}
\affiliation{Department of Physics, University of Chicago, Chicago, Illinois 60637, USA}

\selectlanguage{english}%

\author{Dam Thanh Son}
\email{dtson@uchicago.edu}

\affiliation{Department of Physics, University of Chicago, Chicago, Illinois 60637, USA}

\affiliation{Kadanoff Center for Theoretical Physics, University of Chicago, Chicago, Illinois 60637, USA}

\affiliation{Enrico Fermi Institute and James Franck Institute, University of Chicago, Chicago, Illinois 60637, USA}

\author{Chaolun Wu}
\email{chaolunwu@uchicago.edu}
\affiliation{Kadanoff Center for Theoretical Physics, University of Chicago, Chicago, Illinois 60637, USA}

\begin{abstract}
  We show that for a class of model Hamiltonians for which certain
  trial quantum Hall wavefunctions are exact ground states, there is a
  single spectral density function which controls all two-point
  correlation functions of density, current and stress tensor
  components.  From this we show that the static structure factors
  of these wavefunctions
  behaves at long wavelengths as $s_4 k^4$ where the coefficient $s_4$
  is directly related to the shift: $s_4=(\mathcal S-1)/8$.
\end{abstract}
\maketitle

\emph{Introduction.}---Starting from the work of
Laughlin~\cite{Laughlin:1983fy}, trial wavefunctions have been playing
a very important role in quantum Hall physics.  Some of the most
interesting quantum Hall phases have been predicted theoretically by
the explicit construction of trial wavefunctions, most notably the
Moore-Read (Pfaffian) state~\cite{Moore:1991ks} and the Read-Rezayi
parafermion series~\cite{Read:1998ed}.  A more recent construction
involves the Jack polynomials~\cite{Bernevig:2008rda} and includes an
earlier proposed ``Gaffnian'' state~\cite{Gaffnian}; map
of these states, however, are not expected to correspond to gapped
quantum Hall states.  In most cases, the trial wavefunction is a
ground state of a model Hamiltonian containing only local
interactions~\cite{Haldane:1983xm}.  The study of these wavefunctions
is important for the understanding of the properties of the quantum
Hall phases.

In this paper we will show, among other results, that for a large
class of trial wavefunctions, the low-wavelength asymptotics of the
projected structure factor~\cite{Girvin:1986zz} is determined exactly
by the shift,
\begin{equation}\label{sk}
  s_4 \equiv \lim_{k\to0} \frac{\bar s(k)}{(k\ell_B)^4} = \frac{\mathcal S-1}8
  \,,
\end{equation}
where $\ell_B$ is the magnetic length that we frequently will set to 1.
This equation is already known to be valid for the Laughlin states
with filling fractions $\nu=1/(2n+1)$ where
$s_4=(1-\nu)/(8\nu)$~\cite{Girvin:1986zz} and $\mathcal S=1/\nu$.  In
this paper we show that Eq.~(\ref{sk}) is valid also for the
Moore-Read states, the Read-Rezayi parafermion states, both for bosons
and fermions~\footnote{We consider in this paper only states which in the terminology
  of Ref.~\cite{Read:1998ed} have $q=1$ and $q=2$.}.
Equation (\ref{sk}) was speculated to be valid in
Ref.~\cite{Haldane:2009ke} for all states whose wavefunctions are
constructed from conformal field theory correlators.  The statement
was has not been proven rigorously for any states beyond the Laughlin
states. Instead the \emph{inequality},
\begin{equation}\label{sk-ineq}
  \lim_{k\to0} \frac{\bar s(k)}{k^4} \ge \frac{\mathcal |S-1|}8 \,,
\end{equation}
has been shown to be valid for all gapped ground states on the lowest
Landau level (LLL)~\cite{Haldane:2009ke,Golkar:2013gqa}.
Equation~(\ref{sk}) shows that the trial ground states are truly
special in their respective classes---these are the states that
minimize the structure factor at small $k$.

Our proof of Eq.~(\ref{sk}) is not a direct one, but but relies on
some techniques which reveal some other unusual properties of the
trial states.  First we find the LLL expression for the components of
the stress tensor.  The LLL form of the electromagnetic current has
been found in the
past~\cite{Stone1,Rajaraman:1994iz,RajaramanSondhi:1994}, but the
stress tensor has not been obtained in these works.  We then show
explicitly that the trial ground states in the Read-Rezayi parafermion
series (which includes the Moore-Read state) are annihilated by one
component of the particle number current, as well as all but one
components of of the stress tensor.  Namely,
\begin{equation}
  J_{\bar z}(x) |0\rangle = 0, \quad T_{\bar z\bar z}(x)|0\rangle=
  T_{z\bar z}(x)|0\rangle= 0 ,
\end{equation}
where $|0\rangle$ denotes the ground state.  These equations imply
that all two-point correlation functions involving the density, the
current, and the stress tensor are determined by one single spectral
density.  From this Eq.~(\ref{sk}) follows.

\emph{Action principle for a system on the lowest Landau level}.---In
principle, the form of the stress tensor can be obtained by a
procedure similar to the one followed in
Refs.~\cite{Stone1,Rajaraman:1994iz,RajaramanSondhi:1994} for the
electromagnetic current.  One would develop a perturbation theory in
the inverse cyclotron frequency and pick out the terms that survive
when the cyclotron frequency goes to infinity.  Here we use a much
simpler method based on the Lagrangian formalism.

To derive the form of the stress tensor, we put the system in an
curved metric.  We consider the system of two-dimensional
non-relativistic particles interacting with gauge potential $A_{\mu}$
in curved space \cite{Son1} with $g=\det(g_{ij})$
\begin{multline}\label{S-orig}
  S = \int\! d^3 x\,\sqrt g \Big[ \frac i2
  \left( \psi^\+ D_t\psi - D_t\psi^\+\psi \right)
  -\frac{g^{ij}}{2m} D_i\psi^\+ D_j\psi \\
  + \frac B{2m}\psi^\+\psi
  + \mathcal{L}_{\mathrm{int}} \Big],
\end{multline}
where $\psi$ is the spinless field operator,
$\mathcal{L}_{\mathrm{int}}$ the interacting Lagrangian density.  To
facilitate taking the $m\to0$ limit we have added a magnetic moment
term with the gyromagnetic ratio equal to 2; this term does not affect
the ground state wavefunction in constant magnetic field.  The magnetic
field is $B=\frac{\epsilon^{ij}}{\sqrt g}\dd_i A_j$.  The
covariant derivatives are defined as
\begin{equation}
  D_\mu =\dd_\mu - iA_\mu + i\omega_\mu,
\end{equation}
where the spin connection in term of vielbein is \cite{Son2} 
\begin{equation}
  \omega_t=\frac12\epsilon_{ab}e^{aj}\dd_t e_j^b,\quad
  \omega_i=\frac12\Big(\epsilon_{ab}e^{aj}\dd_i e_j^b
    -\frac{\epsilon^{jk}}{\sqrt g}\dd_j g_{ik}\Big).
\end{equation}
The inclusion of $\omega_\mu$ in the covariant derivative is optional,
but it simplifies the Ward identities used later.
For convenience, we define the complex vielbein vectors
\begin{equation}
  e_i=\frac1{\sqrt2}(e_i^1 - ie_i^2),\quad
  \bar e_i=\frac1{\sqrt2}(e_i^1 + ie_i^2),
\end{equation}
and introduce the following notation for the projection of each vector $X_i$
on the complex vielbeins,
\begin{equation}
  X=e^i X_i ,\quad \bar X=\bar e^i X_i.
\end{equation}
In flat space we can choose $e^a_i=\delta^a_i$, then
$\dd=\sqrt2\dd_z$, $\bar\dd=\sqrt 2\dd_{\bar z}$.  For the lack of
better terminology, we will call $X$ the ``holomorphic'' component and
$\bar X$ the ``antiholomorphic'' component of $X_i$, without
committing to any particular dependence of $X$ and $\bar X$ on the
spatial coordinates.

The action~(\ref{S-orig}) contains term that are singular in the
$m\to0$ limit.  To have a smooth $m\to0$ limit, we notice that $D_i
D^i+B=D\bar D$ and use the Hubbard-Stratonovich transformation to
rewrite the action in the form
\begin{multline}\label{eq:Lag}
  S =\int\! d^3x\,\sqrt g \Bigl[\frac i2 \left(
    \psi^\+ D_t\psi -D_t\psi^\+\psi\right) \\
    - D\psi^\+\chi - \chi^\+ \bar D\psi
    +m\chi^\+\chi+\mathcal{L}_{\mathrm{int}}\Bigr].
\end{multline}
In the LLL limit $m\to0$, $\chi$ and $\chi^\+$ play the role of
Lagrange multipliers enforcing the constraint
\begin{equation}\label{eq:constr}
  \bar D \psi =0,\qquad D\psi^\+ = 0.
\end{equation}
This is simply the lowest Landau level constraint, which in flat space
becomes $D_{\bar z}\psi=0$.  In the symmetric gauge $A_x=-\frac12By$,
$A_y=\frac12Bx$ the constraint implies that $\psi$ is proportional to
a linear combination of $z^n e^{-B|z|^2/4}$.  The time evolution follows the
equation of motion
\begin{equation}
  i\dd_t\psi + D\chi + A_t\psi
  +\frac{\delta L_{\mathrm{int}}}{\delta\psi^\+}  =0,\label{eq:masterchi}
\end{equation}
where $L_{\mathrm{int}}=\int\! d{\vec x}\,\mathcal{L}_{\mathrm{int}}$,
and its complex conjugate.  In Eq.~(\ref{eq:masterchi}) the Lagrange
multiplier $\chi$ is such that the constraint~(\ref{eq:constr}) is
maintained at all times.  We shall determine $\chi$ from this
condition later.

We will mostly consider in this paper $\mathcal L_{\rm int}$ that
contains only local interactions, i.e., interactions that are given by
a product of $\psi$, $\psi^\+$ and their derivatives at the same
point.  In the first-quantized language such an interaction
corresponds to a many-body potential in the form of a product of delta
functions and their derivatives.  Upon projection onto the LLL, such
interactions become the pseudopotential interactions.  Due to the LLL
constraint, any such potential can be written as $\mathcal L_{\rm
  int}[\psi^\+,\psi,\bar D^n\psi^\+, D^n\psi]$.

The form of the interaction Lagrangian is chosen so that the ground
state is the trial wave function under consideration.  in this paper
the interaction term that lead to a well-known series of trial
wavefunctions: the Read-Rezayi parafermion series.  For bosons, our
interaction Lagrangian has the form
\begin{equation}\label{Lintb}
  \mathcal L_{\rm int} = -\lambda (\psi^\+)^k \psi^k,
\end{equation}
and for fermions,
\begin{equation}\label{Lintf}
  \mathcal L_{\rm int} = - \lambda\bigl|\psi D\psi D^2\psi\cdots
    D^{k-1}\psi\bigr|^2.
\end{equation}
These interactions place an energy penalty on $k$ particles coming in
with minimal angular momentum.  For case $k=2$ corresponds to the
$\nu=1/2$ bosonic and $\nu=1/3$ fermionic Laughlin states; $k=3$
corresponds to the Moore-Read states, and $k>3$ to the parafermion
states.

\emph{Currents and stress tensor.}---We define the charge current
$J^{\mu}$ and the stress tensor $T^{ij}$ from the variation of the
action with respect to the background fields,
\begin{equation}
  \delta S=\int\! d^{3}x\,\sqrt{g}\,\Bigl(J^{\mu}\delta A_{\mu}
    +\frac12T^{ij}\delta g_{ij}\Bigr) .
\end{equation}
where the action is given by Eq.~(\ref{eq:Lag}).  For convenience, in
addition to the holomorphic and antiholomorphic components of the
current $J=J^{i}e_{i}$, $\bar J=J^i\bar e_i$, we introduce
\begin{align}
  T &=T^{ij}e_ie_j =2T_{zz},\quad
  \tilde T &= T^{ij}\bar e_i\bar e_j = 2T_{\bar z\bar z},\\
  T^{\rm tr} &= T^{ij}e_i \bar e_j = 2 T_{z\bar z}.
\end{align}
We call $T$ and $\tilde T$ the holomorphic and antiholomorphic stress components; $T^{\rm tr}$ is simply the trace of of the stress tensor.
By varying the Lagrangian, we find
\begin{align}
  \rho\equiv J^0 & =\psi^\+\psi ,\quad
  J =i\chi^\+ \psi ,\quad
  \bar J = -i\psi^\+\chi.
\end{align}
Note that the $\mathcal L_{\rm int}$, even in the fermion case, does
not depend on the gauge potential $A_\mu$ and hence does not
contribute to the current.

For the stress tensor we find
\begin{subequations}\label{TTbar}
\begin{align}
  \tilde T & = \bar D \psi^\+\chi - \bar\dd (\psi^\+\chi)
  +\bar e_i\frac{\delta L_{\mathrm{int}}}{\delta e_i} \,,\label{eq:T}\\
  T & =\chi^\+D\psi - \dd(\chi^\+\psi)
  +e_i\frac{\delta L_{\mathrm{int}}}{\delta \bar e_i}\,,\label{eq:barbarT}\\
  T^{\rm tr} & = 2\mathcal{L}_{\mathrm{int}}
 + e_{i}\frac{\delta L_{\mathrm{int}}}{\delta e_{i}}+\bar{e}_{i}\frac{\delta L_{\mathrm{int}}}{\delta\bar{e}_{i}}-\frac{\delta L_{\mathrm{int}}}{\delta\psi}\psi-\psi^{\dagger}\frac{\delta L_{\mathrm{int}}}{\delta\psi^{\dagger}}.
\end{align}
\end{subequations}
where $\chi$ is given in (\ref{eq:chi_LLL}).  Now the bosonic $L_{\rm
  int}$ does not contain any $e^i$ and the last terms in
Eqs.~(\ref{eq:T}) and (\ref{eq:barbarT}) equal zero.  For the fermionic
interaction~(\ref{Lintf}), the operator $\bar e_i\delta/\delta e_i$
converts one $D$ into a $\bar D$, which can be pushed to act on $\psi$
by using $[\bar D,\,D]=-B$.  Due to the constraints $\bar D\psi=0$ the
result will be a product of $k$ derivatives of $\psi$ but two of the
derivatives will have the same power.  Fermion statistics then implies that the result is zero.  Thus we find that $L_{\rm int}$ does not contribute to the traceless components of the stress tensor and
\begin{subequations}\label{TTbar-final}
\begin{align}
  \tilde T & = \bar D \psi^\+\chi - \bar\dd (\psi^\+\chi),\label{eq:T-final}\\
  T & =\chi^\+D\psi - \dd(\chi^\+\psi).\label{eq:barbarT-final}
\end{align}
\end{subequations}

\emph{Special properties of trial ground state.}---We now show that
the trial ground states satisfy the following properties:
\begin{align}\label{fund_prop}
  \bar J(x) |0\rangle &= 0,\\
  \bar T(x) |0\rangle &=0,\\
  T^{\rm tr}(x) |0\rangle &=0.
\end{align}
To show the first two relations it is sufficient to show that $\chi$
annihilate the ground state.  For that, we need to solve the
equation~(\ref{eq:masterchi}) and find $\chi$.  First we act the
operator $\bar D$ on Eq.~(\ref{eq:masterchi}) and use
Eq.~(\ref{eq:constr}) to get
\begin{equation}\label{DDchiDF}
  \bar DD \chi + \bar D\mathcal F = 0, \quad
  \mathcal F= A_t\psi + \frac{\delta L_{\rm int}}
  {\delta\psi^\+} \,.
  \end{equation}
Now we note that we can express the projection
of any function $\mathcal{F}(\vec{x})$ onto the LLL as an expansion
over derivatives,
\begin{equation}
\mathcal{F}_{\mathrm{L}}(\vec{x})=\mathcal{P}_{\mathrm{LLL}}\mathcal{F}(\vec{x})=\sum_{n=0}^{\infty}\frac{1}{n!B^{n}}D^n\bar D^{n}\mathcal{F}(\vec{x}).\label{eq:proj}
\end{equation}
In particular, one can check that this is consistent with
\begin{equation}
  \bar D\mathcal{F}_{\mathrm{L}}(\vec{x})=0,
  \qquad\left(\mathcal{P}_{\mathrm{LLL}}\right)^{2}=\mathcal{P}_{\mathrm{LLL}}.
\end{equation}
Replacing in Eq.~(\ref{DDchiDF}) $\mathcal F$ by $\mathcal F-\mathcal F_{\rm L}$, then use the series expansion~(\ref{eq:proj}) for $\mathcal F_{rm L}$, one finally finds $\chi$ as a series
\begin{equation}
  \chi  =\sum_{n=1}^\infty\frac1{n!B^{n}} D^{n-1}\bar D^n
  \left(A_{t}\psi+\frac{\delta L_{\rm int}}{\delta\psi^\+}\right).
  \label{eq:chi_LLL}
\end{equation}
In particular, if one considers noninteracting electrons, $L_{\rm
  int}=0$ and insert $\chi$ into the expression for the current, one
reproduces the expression previously obtained by Mart\'inez and
Stone~\cite{Stone1}.

We now set $A_t=0$ and inspect the operator $\chi$.  In both the
bosonic and the fermionic cases, taking the variation over $\psi^\+$
leaves the strings of annihilation operators, $\psi^k$ and $\psi D\psi
D^2\psi\cdots$ intact on the right of $L_{\rm int}$.  But the trial
wavefunction is annihilated by this exact string of annihilation
operators,
\begin{align}
  \psi^k(x) | 0 \rangle_{\rm bosonic} &= 0,\\
  \psi D\psi D^2\psi \cdots D^{k-1}\psi |0\rangle_{\rm fermionic} &= 0.
\end{align}
We thus conclude that $\chi(x)$ annihilates the ground state, and from
the explicit forms for the current and stress tensor, one concludes
that the antiholomorphic components $\bar J$ and $\tilde T$
annihilates the ground state.

Similar calculations also show that the trace of the stress tensor
$T^{\rm tr}$ also annihilates the ground state.

The properties~(\ref{fund_prop}) are properties specific for the trial
ground states and the Hamiltonian for which these ground states are
exact zero-energy states.  They are not valid for generic interaction,
for example the Coulomb interactions.

\emph{Ward identities.}---In flat spacetime we have the conservation
laws for the particle number and momentum, which is in our notation are
\begin{align}
  \dd_t \rho + \bar{\partial}J+\partial\bar{J} & =0,\label{eq:CurrentConservation}\\
  \dd \tilde T+\bar\dd T^{\rm tr} &=-iB\bar J,\label{eq:StressConservation}\\
  \dd T^{\rm tr}+\bar\dd T & = +iBJ.
\end{align}
The last two equations are simply the force balance equations, since
we are working in the limit $m\to0$ where there is no inertia.  We now
sandwich these equations between the ground state $|0\rangle$ and an
arbitrarily chosen state $\langle n|$, assuming that the latter is a
state with zero particle number and carries energy $E_n$ and momentum
${\bf P}_n$.  Since $\bar J$, $\tilde T$ and $T^{\rm tr}$ annihilate
the ground state, thee Ward identities imply direct proportionality
between the matrix elements of the operators $\rho$, $J$ and $T$,
\begin{align}
  \langle n | \rho |0 \rangle &= -\frac{(P_n^x+iP_n^y)^2}{BE_n}
  \langle n | T_{zz} |0 \rangle ,\\
  \langle n | J_z |0 \rangle &= -\frac{P_n^x+iP_n^y}B
  \langle n | T_{zz} |0 \rangle .
\end{align}
This means that the three operators $\rho$, $J$ and $T$ create the
same set of states at nonzero momentum, only with different matrix
elements.  For example, if a magneto-roton~\cite{Girvin:1986zz} exists
it can be created equally well by all three operators.  This does not
apply to states with zero momentum (including the magneto-roton if it
exists at zero momentum); these states cannot be created by acting
$\rho$ or $J$ on the ground state, but may be created by the operator
$T$.

If one introduces the spectral densities of the density and the holomorphic component of the stress tensor,
\begin{align}
  S(\omega, k) &= \frac1N \sum_n |\langle n| \rho({\bf k})|0\rangle|^2
  \delta(\omega-E_n), \\
  \rho_T(\omega, k) &= \frac1N\sum_n|\langle n| T_{zz}({\bf k})|0\rangle|^2
  \delta(\omega-E_n),
\end{align}
then
\begin{equation}
  \omega^2 S(\omega, k) = \frac{k^4}{B^2} \rho_T(\omega, k).
\end{equation}

The static structure factor can be expressed as
\begin{equation}\label{eq:SumRule_Sk}
  S(k) = \int\limits_0^\infty\!d\omega\, S(\omega,k) = k^4\!
  \int\limits_0^\infty\!\frac{d\omega}{\omega^2} \rho_T(\omega,k).
\end{equation}
So at $k\to0$ the static structure factor $S(k)$ is proportional to
$k^4$.  Since by definition our spectral density $S(\omega,k)$ does
not include the cyclotron mode, $S(k)$ is actually the projected
structure factor $\bar s(k)$, which is known to be $O(k^4)$ at small $k$.

On the other hand, the retarded Green function of two components of
the stress tensor $T^{ij}$ and $T^{kl}$ can be decomposed as~\cite{Tokatly0706,Tokatly0812}: 
\begin{equation}
  G_{\mathrm{R}}^{T^{ij},T^{kl}}(\omega,{\vec 0})=K(\omega)\mathcal{I}_{\mathrm{B}}^{ijkl}+\mu(\omega)\mathcal{I}_{\mathrm{S}}^{ijkl}-i\omega\eta_{\mathrm{H}}(\omega)\mathcal{I}_{\mathrm{H}}^{ijkl},\label{eq:RetardedTT_Expansion}
\end{equation}
where $K(\omega)$, $\mu(\omega)$ and $\eta_{\mathrm{H}}(\omega)$
are the frequency-dependent bulk modulus,
shear modulus and Hall viscosity, respectively, and
\begin{align*}
\mathcal{I}_{\mathrm{B}}^{ijkl} & =\delta^{ij}\delta^{kl},\\
\mathcal{I}_{\mathrm{S}}^{ijkl} & =\delta^{ik}\delta^{jl}+\delta^{il}\delta^{jk}-\delta^{ij}\delta^{kl},\\
\mathcal{I}_{\mathrm{H}}^{ijkl} & =\frac{1}{2}\left(\delta^{ik}\epsilon^{jl}+\delta^{il}\epsilon^{jk}+\delta^{jk}\epsilon^{il}+\delta^{jl}\epsilon^{ik}\right),
\end{align*}
For the three independent response functions on ground states with
$\tilde T|0\rangle=0$, the analytic structure of the retarded 2-point
function implies the following sum rule,
\begin{align}
  \int\limits_0^\infty\!\frac{d\omega}{\omega^2}\,
  \rho_T(\omega) & =\frac{\eta_{\mathrm{H}}(0)-\eta_{\mathrm{H}}(\infty)}{2\rho},
\end{align}
Compared with (\ref{eq:SumRule_Sk}), we have for our ground state
\begin{equation}
  \lim_{k\to0}\frac{S(k)}{k^4} =\frac{\eta_{\mathrm{H}}(0)-\eta_{\mathrm{H}}(\infty)}{2\rho},
\end{equation}
Now let us recall that the Hall viscosity (at zero frequency) of a
gapped quantum Hall state is equal to $\eta_{\rm H}(0)=\rho{\cal S}/4$
where $\mathcal S$ is the shift of the state~\cite{Read:2008rn}.
At frequencies much
higher than the energy scale set by the interaction, interactions do
not play any roles and the Hall viscosity is given by the same formula
as in the integer quantum Hall state of the lowest Landau level,
$\eta_{\rm H}(\infty)=\rho/4$.  Therefore the previous equation can be
  written as
\begin{equation}
  \lim_{k\to0}\frac{S(k)}{k^4} = \frac{\mathcal S-1}8
\end{equation}
The fact that this relationship is valid for the Laughlin wavefunction
is well known.  What we have shown is that this relationship is valid
for a much wider class of ground states.  In fact, one can show that
the relationship is valid whenever the ground state is annihilated by
the uniform component of the antiholomorphic component of the stress tensor,
\begin{equation}
  \int\!d{\bf x}\, \tilde T({\bf x}) |0\rangle = 0
\end{equation}
The trial ground states and their corresponding interaction
Hamiltonian satisfy a stronger constraint $\tilde T(x)|0\rangle=0$.

\emph{Conclusion.}---Our result, Eq.~(\ref{sk}) shows that there is a
special class of quantum Hall wavefunctions that saturate the
inequality~(\ref{sk-ineq}).  In these wavefunctions, the leading $k^4$
behavior of the static structure factor is related to the shift.  It
is also clear that the relationship $s_4=(\mathcal S-1)/8$ cannot be
valid for all gapped quantum Hall states.  For examples, for states
that have $\mathcal S<1$, for example the $\nu=2/3$ state ($\mathcal
S=0$) or the anti-Pfaffian state~\cite{Levin:2007,SSLee:2007}
($\mathcal S=-1$), $(\mathcal S-1)/8<0$ while, due to the positivity
of the dynamic structure factor $S(\omega,k)$, the coefficient $s_4$
has to be positive.


It is known that many of the trial wave functions have large overlaps
with the true ground state of the Hamiltonian with Coulomb
interaction.  It is interesting to see if the numerical value of the
$k^4$ coefficient in the structure factor is close to the value
$(\mathcal S-1)/8$ achieved by the trial wavefunctions.

The authors thank Nick Read and Paul Wiegmann for useful discussions.
We acknowledge a communication by Nick Read~\cite{Read_private} in
which he outlines a proof of the equality~(\ref{sk}) using the strain
generators introduced in Ref.~\cite{Bradlyn:2012ea}.  Our proof uses
only the local stress tensor and hence avoids the issue of the edge
modes.  We thank Paul Wiegmann for sharing with us an upcoming
paper~\cite{CLW} where the Eq.~(\ref{sk}) is proven directly from the
wavefunction without using properties of the excited states.  This
work is supported, in part, by DOE grant DE-FG02-13ER41958 and a
Simons Investigator grant from the Simons Foundation.


\begin{thebibliography}{99}

\bibitem{Laughlin:1983fy} 
  R.~B.~Laughlin,
  \emph{Anomalous quantum Hall effect: An incompressible quantum fluid with fractionally charged excitations,}
  Phys.\ Rev.\ Lett.\  {\bf 50}, 1395 (1983).

\bibitem{Moore:1991ks} 
  G.~W.~Moore and N.~Read,
  \emph{Nonabelions in the fractional quantum Hall effect,}
  Nucl.\ Phys.\ B {\bf 360}, 362 (1991).

\bibitem{Read:1998ed} 
  N.~Read and E.~Rezayi,
  \emph{Beyond paired quantum Hall states: Parafermions and incompressible states in the first excited Landau level,}
  Phys.\ Rev.\ B {\bf 59}, 8084 (1999)
  [cond-mat/9809384].

 \bibitem{Bernevig:2008rda} 
  B.~Bernevig and F.~Haldane,
  \emph{Model Fractional Quantum Hall States and Jack Polynomials,}
  Phys.\ Rev.\ Lett.\  {\bf 100}, 246802 (2008)
  [arXiv:0707.3637].

\bibitem{Gaffnian}
  S.~H.~Simon, E.~H.~Rezayi, N.~R.~Cooper, and I.~Berdnikov,
  \emph{Construction of a paired wave function for spinless electrons at
  filling fraction $\nu=2/5$,}
  Phys. Rev. B {\bf 75}, 075317 (2007)
  [cond-mat/0608376].

\bibitem{Haldane:1983xm} 
  F.~D.~M.~Haldane,
  \emph{Fractional quantization of the Hall effect: A Hierarchy of incompressible quantum fluid states,}
  Phys.\ Rev.\ Lett.\  {\bf 51}, 605 (1983).

\bibitem{Girvin:1986zz} 
  S.~M.~Girvin, A.~H.~MacDonald, and P.~M.~Platzman,
  \emph{Magneto-roton theory of collective excitations in the fractional quantum Hall effect,}
  Phys.\ Rev.\ B {\bf 33}, 2481 (1986).
  
\bibitem{Haldane:2009ke} 
  F.~D.~M.~Haldane,
  \emph{Hall viscosity and intrinsic metric of incompressible fractional Hall fluids,}
  arXiv:0906.1854.

\bibitem{Golkar:2013gqa} 
  S.~Golkar, D.~X.~Nguyen and D.~T.~Son,
  \emph{Spectral Sum Rules and Magneto-Roton as Emergent Graviton in Fractional Quantum Hall Effect,}
  arXiv:1309.2638.
  
\bibitem{Stone1} J. Mart\'inez and M. Stone,
  \emph{Current Operators in the Lowest Landau Level,}
  Int. J. Mod. Phys. B \textbf{7}, 4389 (1993).

\bibitem{Rajaraman:1994iz} 
  R.~Rajaraman,
  \emph{Currents in the lowest Landau level field theory with e e interactions,}
  Int.\ J.\ Mod.\ Phys.\ B {\bf 8}, 777 (1994).

\bibitem{RajaramanSondhi:1994}
  R.~Rajaraman and S.~L.~Sondhi,
  \emph{Landau level mixing and solenoidal terms in lowest Landau level
  currents,}
  Mod.\ Phys.\ Lett. B {\bf 8}, 1065 (1994)
  [cond-mat/9406020].
  

\bibitem{Son1} Michael Geracie, Dam Thanh Son, Chaolun Wu and Shao-Feng
Wu, \emph{Spacetime Symmetries of the Quantum Hall Effect,}
arXiv:1407.1252.

\bibitem{Son2} D. T. Son, \emph{Newton-Cartan Geometry and the
Quantum Hall Effect,} arXiv:1306.0638.


\bibitem{Tokatly0706}  I.~V.~Tokatly and G.~Vignale, \emph{Lorentz shear modulus of a two-dimensional electron gas at high magnetic field,} Phys.\ Rev.\ B {\bf 76}, 161305 (2007) [arXiv:0706.2454].

\bibitem{Tokatly0812} I.~V.~Tokatly and G.~Vignale, \emph{Lorentz shear modulus of fractional quantum Hall states,} J.\ Phys.:\ Condens.\ Matter {\bf 21}, 275603 (2009) [arXiv:0812.4331].

\bibitem{Levin:2007}
M.~Levin, B.~I.~Halperin, and B.~Rosenow,
\emph{Particle-Hole Symmetry and the Pfaffian State,}
Phys.\ Rev.\ Lett. {\bf 99}, 236806 (2007)
[arXiv:0707.0483].

\bibitem{SSLee:2007}
S.-S.~Lee, S.~Ryu, C.~Nayak, and M.~P.~A.~Fisher,
\emph{Particle-Hole Symmetry and the $\nu=5/2$ Quantum Hall State,}
Phys.\ Rev.\ Lett.\ {\bf 99}, 236807 (2007)
[arXiv:0707.0478].


  
\bibitem{Read:2008rn} 
  N.~Read,
  \emph{Non-Abelian adiabatic statistics and Hall viscosity in quantum Hall states and $p_x + ip_y$ paired superfluids,}
  Phys.\ Rev.\ B {\bf 79}, 045308 (2009)
  [arXiv:0805.2507].
  
\bibitem{Read_private}
  N.~Read, private communication.
  
\bibitem{Bradlyn:2012ea} 
  B.~Bradlyn, M.~Goldstein and N.~Read,
  \emph{Kubo formulas for viscosity: Hall viscosity, Ward identities, and the relation with conductivity,}
  Phys.\ Rev.\ B {\bf 86}, 245309 (2012)
  [arXiv:1207.7021].

\bibitem{CLW}
  T.~Can, M.~Laskin, and P.~Wiegmann,
  \emph{Fractional Quantum Hall Effect: Gravitational Anomaly and Kinetic Coefficients,} to be published.
  

\end{thebibliography}
\end{document}